\documentclass[aps,prl,twocolumn,superscriptaddress,nofootinbib]{revtex4}

\usepackage{lineno,hyperref}

\usepackage[T1]{fontenc}
\usepackage[latin9]{inputenc}
\usepackage{amsmath}

\usepackage{graphicx}
\usepackage{comment}
\usepackage{ulem}

\usepackage{color}
\usepackage{graphicx}
\usepackage{multirow}
\usepackage{amsmath,bm}

\newcommand{\XeXe}{$^{129}$Xe+$^{129}$Xe}
\newcommand{\mean}[1]{\langle#1\rangle}

\newcommand {\ivishnu}{iEBE-VISHNU}
\newcommand {\vishnu}{VISH2+1}
\newcommand {\trento}{TRENTo}
\newcommand {\Xe}{$^{129}$Xe}
\newcommand {\vnpt}{$v_2^2-[p_T]$}

\newcommand {\dperp}{d_\perp}

\begin{document}


\title{Exploring the Nuclear Shape Phase Transition in Ultra-Relativistic \XeXe\  Collisions at the LHC} 

\author{Shujun Zhao}
\affiliation{School of Physics, Peking University, Beijing 100871, China}
\affiliation{School of Science, Huzhou University, Huzhou, Zhejiang 313000, China}
\author{Hao-jie Xu}
\affiliation{School of Science, Huzhou University, Huzhou, Zhejiang 313000, China}
\affiliation{Strong-Coupling Physics International Research Laboratory (SPiRL), Huzhou University, Huzhou, Zhejiang 313000, China.}
\author{You Zhou}
\affiliation{Niels Bohr Institute, University of Copenhagen, Blegdamsvej 17, 2100 Copenhagen, Denmark}
\author{Yu-Xin Liu}
\affiliation{School of Physics, Peking University, Beijing 100871, China}
\affiliation{Collaborative Innovation Center of Quantum Matter, Beijing 100871, China}
\affiliation{Center for High Energy Physics, Peking University, Beijing 100871, China}
\author{Huichao Song}
\affiliation{School of Physics, Peking University, Beijing 100871, China}
\affiliation{Collaborative Innovation Center of Quantum Matter, Beijing 100871, China}
\affiliation{Center for High Energy Physics, Peking University, Beijing 100871, China}




\date{\today}

\begin{abstract}
The shape phase transition for certain isotope or isotone chains, associated with the quantum phase transition of finite nuclei, is an intriguing phenomenon in nuclear physics. A notable case is the Xe isotope chain, where the structure transits from a $\gamma$-soft rotor to a spherical vibrator, with the second-order shape phase transition occurring in the vicinity of $^{128-130}$Xe. In this letter, we focus on investigating the $\gamma$-soft deformation of \Xe\ associated with the second-order shape phase transition by constructing novel correlators for ultra-relativistic \XeXe\ collisions. In particular, our iEBE-VISHNU model calculations show that the correlation between elliptic flow $v_{2}$ and mean transverse momentum $[p_{T}]$, denoted as $\rho_{2}$, as well as the $[p_{T}]$ fluctuation $\Gamma_{p_T}$, which were previously used to claim the evidence of the rigid triaxial deformation of \Xe, can also be well explained by the $\gamma$-soft deformation of \Xe. We further propose two novel correlators $\rho_{4,2}$ and $\rho_{2,4}$, which carry non-trivial higher-order correlations and show unique capabilities to distinguish between the $\gamma$-soft and the rigid triaxial deformation of \Xe\ in \XeXe\ collisions at the LHC. The present study provides a novel way to explore the second-order shape phase transition of finite nuclei with ultra-relativistic heavy ion collisions.
\end{abstract}

\maketitle

\paragraph{Introduction.}
The phase transition has been studied extensively in various research areas of physics. In high-energy nuclear physics, the heavy ion program at the Relativistic Heavy Ion Collider (RHIC) and the Large Hadron Collider (LHC) aims to explore the QCD phase transition and to study the quark-gluon plasma (QGP)~\cite{STAR:2005gfr, PHENIX:2004vcz, BRAHMS:2004adc, PHOBOS:2004zne, ALICE:2022wpn, Gyulassy:2004zy, Muller:2012zq,  Shuryak:2014zxa}. 
In nuclear structure physics, the concept of the shape phase transition of finite nuclei has been developed to explain the rapid structural change along certain isotope or isotone chains, associated with the dynamic interplay between the spherical-driving pairing interaction and the deformation-driving proton-neutron interaction~\cite{Casten:2009zz, Casten_2007, Casten:2007zz, RevModPhys.82.2155, CASTEN2009183}.
A deeper understanding of the shape phase transition is crucial to unravel the collective properties residing in atomic nuclei. In this paper, we make the first attempt to probe the possible second-order shape phase transition for the Xe isotopes in the ultra-relativistic \XeXe\ collisions at the LHC.

In Refs.~\cite{Iachello:2000ye, Iachello:2001ph}, the critical point symmetries such as E(5) and X(5) have been proposed to describe and classify the shape phase transition of finite nuclei.
In the framework of the Interacting Boson Model (IBM)~\cite{Iachello_Arima_1987}, the Xe isotopes undergo a shape phase transition from a $\gamma$-soft rotor to a spherical vibrator~\cite{Casten:1985lxb, Puddu:1980mnv, Casten:1985ydp}, with the critical point described by the $E(5)$ symmetry~\cite{Iachello:2001ph}.  
Experimentally, the measured energy spectroscopy of $^{128}$Xe agrees well with the predictions of the $E(5)$ symmetry, including the normalized transition strengths and the branching ratios, as well as the energy ratios between different energy levels~\cite{Clark:2004xb}. Theoretically, several microscopic approaches have been used to study the $E(5)$ symmetry for the Xe isotopes~\cite{Li:2010qu, Nomura:2017ilh, Fossion:2006xg, Rodriguez-Guzman:2007fla, Robledo:2008zz}, which suggests that a critical point of the second-order shape phase transition lies in the vicinity of $^{128-130}$Xe, associated with a $\gamma$-soft and  $\beta$-soft deformation. It is also worth mentioning that Ref~\cite{Bonatsos:2005df, Buganu:2015dxa} has proposed a $Z(4)$ symmetry to describe the spectra and $B(E2)$ rates for $^{128, 130, 132}$Xe, where the Bohr Hamiltonian is assumed to be $\beta$-flat with the $\gamma$ frozen at $30^\circ$.

The \XeXe\ collisions at the LHC~\cite{ALICE:2018lao, ALICE:2018cpu, ALICE:2021gxt,  ALICE:2022xip, ALICE:2018hza}, originally aimed at studying the system-size effects of the QGP, can provide a novel way to identify the $\gamma$-soft and $\gamma$-frozen deformation of \Xe\ and to probe the possible second-order shape phase transition of the Xe isotopes. At ultra-relativistic collision energies, the deformation of the colliding \Xe\ is directly imprinted into the initial stage of the created QGP,  followed by a subsequent hydrodynamic evolution for the bulk matter. The advantage stems from the well-understood hydrodynamic response~\cite{Gale:2013da, Heinz:2013th, Song:2017wtw}, which correlates the final-state momentum distributions of the hadrons with the initial-state spatial distribution of the QGP. As a result, the shape (or deformation) of the colliding nuclei could leave messages on the anisotropic momentum distribution of the final hadrons~\cite{Zhang:2021kxj, Ryssens:2023fkv, Nijs:2021kvn, Zhao:2022uhl, Xu:2021uar, Lu:2023fqd, STAR:2024eky}, characterized by the anisotropic flow $v_n$, the mean transverse momentum $\langle[p_T]\rangle$, the correlations between them, and so on.

The correlation between $v_2^2$ and $[p_T]$, denoted as $\rho_{2}$, was investigated using initial-stage calculations for \XeXe\ collisions at the LHC~\cite{Bally:2021qys}. The study claimed that the \Xe\ nucleus exhibits a rigid triaxial deformation with $\gamma = 27^\circ$. However, in this letter, we argue that the measured $\rho_2$ results from \XeXe\ collisions are insufficient to conclusively demonstrate rigid triaxial deformation in \Xe. Instead, these results can also be consistently explained by assuming a $\gamma$-soft deformation for \Xe. Furthermore, we propose new higher-order correlations that can effectively distinguish between $\gamma$-soft and rigid triaxial deformations of colliding nuclei. This provides a novel way to probe the potential second-order shape phase transitions for certain isotope or isotone chains in relativistic heavy ion collisions. 

\paragraph{Model and setups.}
In this paper, we implement the state-of-the-art \ivishnu\ model~\cite{Shen:2014vra, Song:2010aq} to simulate the dynamic evolution of ultra-relativistic \XeXe\ collisions and study the deformation of the colliding nuclei \Xe.  
In \ivishnu, the initial condition is generated by a parameterized \trento\  model~\cite{Moreland:2014oya}, 
and the QGP evolution is simulated by the (2+1)-dimensional relativistic viscous hydrodynamics \vishnu~\cite{Heinz:2005bw, Song:2007ux, Song:2007fn} with the s95-PCE equation of state~\cite{Huovinen:2009yb, Shen:2010uy} as an input. The QGP fireball hadronizes near $T_C$, and the fluid cells are converted into hadrons with the Cooper-Frye descriptions~\cite{Song:2010aq}. Finally, the produced hadrons are fed into the UrQMD hadron cascade model ~\cite{Bass:1998ca, Bleicher:1999xi} to describe the subsequent hadronic evolution, scatterings and decays.

For the deformed colliding nuclei \Xe,  we sample the position of its nucleons with the following Woods-Saxon distribution in \trento:
\begin{align}
\begin{split}
\rho(r) &= \frac{\rho_0}{1+\exp(\frac{r-R(\theta,\phi)}{a_0})},\\
R(\theta,\phi) &= R_0(1+\beta_2[\cos\gamma Y_{2,0}(\theta,\phi) + \sin\gamma Y_{2,2}(\theta,\phi)]).
\end{split}
\end{align}
The definition of $\gamma$ here is the same as the one in ref.~\cite{Jia:2021qyu, Bally:2021qys}, with the real spherical harmonics defined by $Y_{20}=Y_2^0$, $Y_{22}=\frac{1}{\sqrt{2}}(Y_2^2+Y_2^{-2})$. $R_0$ and $a_0$ are the half-width radii and nuclear diffusions, which are set to be $R_0=5.36$ fm and $a_0=0.59$ fm~\cite{Loizides:2014vua}.  The deformation of \Xe\  is characterized by the quadrupole deformation parameter $\beta_2$  and the triaxial deformation parameter $\gamma$  in $R(\theta,\phi)$. 
When \Xe\ is close to the second-order shape phase transition of the Xe isotope chain, both $\beta_2$ and $\gamma$ fluctuate over a wide range~\cite{Iachello:2000ye, Iachello:2001ph}. Ref.~\cite{Dimri:2023wup} has pointed out that such $\beta_2$ fluctuations can be well constrained by the measured lower order flow harmonics $v_2\{2\}$ and $v_2\{4\}$ in relativistic heavy ion collisions. We therefore focus on the construction of sensitive observables for the $\gamma$-soft structure of $^{129}$Xe and use a frozen $\beta_2$ deformation in our model simulations for \XeXe\ collisions.
Considering that the predicted $\beta_2$ for $^{128,130}$Xe ranges from $0.169$ to $0.192$~\cite{Raman:2001nnq, Elekes:2015qje}, we set the $\beta_2$ of \Xe\ as 0.17. 
To model the $\gamma$-soft effects in the events of heavy ion collisions, we sample the fluctuating $\gamma$ from a flat distribution with the range $0^\circ\le\gamma\le60^\circ$. For comparison, we perform calculations for the case of a triaxially deformed \Xe\ with a frozen $\gamma$ set to $\gamma=30^\circ$.

To describe the $\rho_2$ measurements of  \XeXe\ collisions, we use a smaller nucleon width $w=0.5$ fm for the initial state model \trento\ as used in~\cite{Giacalone:2021clp}. 
The remaining parameters in \trento\, the QGP shear and bulk viscosity and other parameters in \ivishnu\ are then tuned based on the results of the Bayesian analysis~\cite{Moreland:2018gsh}, which can describe most of the bulk observables measured in the experiments~\cite{ALICE:2016ccg,ALICE:2018rtz,ALICE:2018lao, ALICE:2018cpu, ALICE:2021gxt, ALICE:2018hza, ALICE:2023tej, ATLAS:2022dov}, including the mean multiplicity $\langle N_{\text{ch}}\rangle$, the elliptic flows with $2$-particle correlation $v_2\{2\}$ and $4$-particle correlation $v_2\{4\}$, and the mean transverse momentum $\langle [p_T]\rangle$.

\begin{figure*}[!hbt]
        \begin{centering}
            \begin{minipage}[t]{0.32\textwidth}
                \includegraphics[scale=0.29]{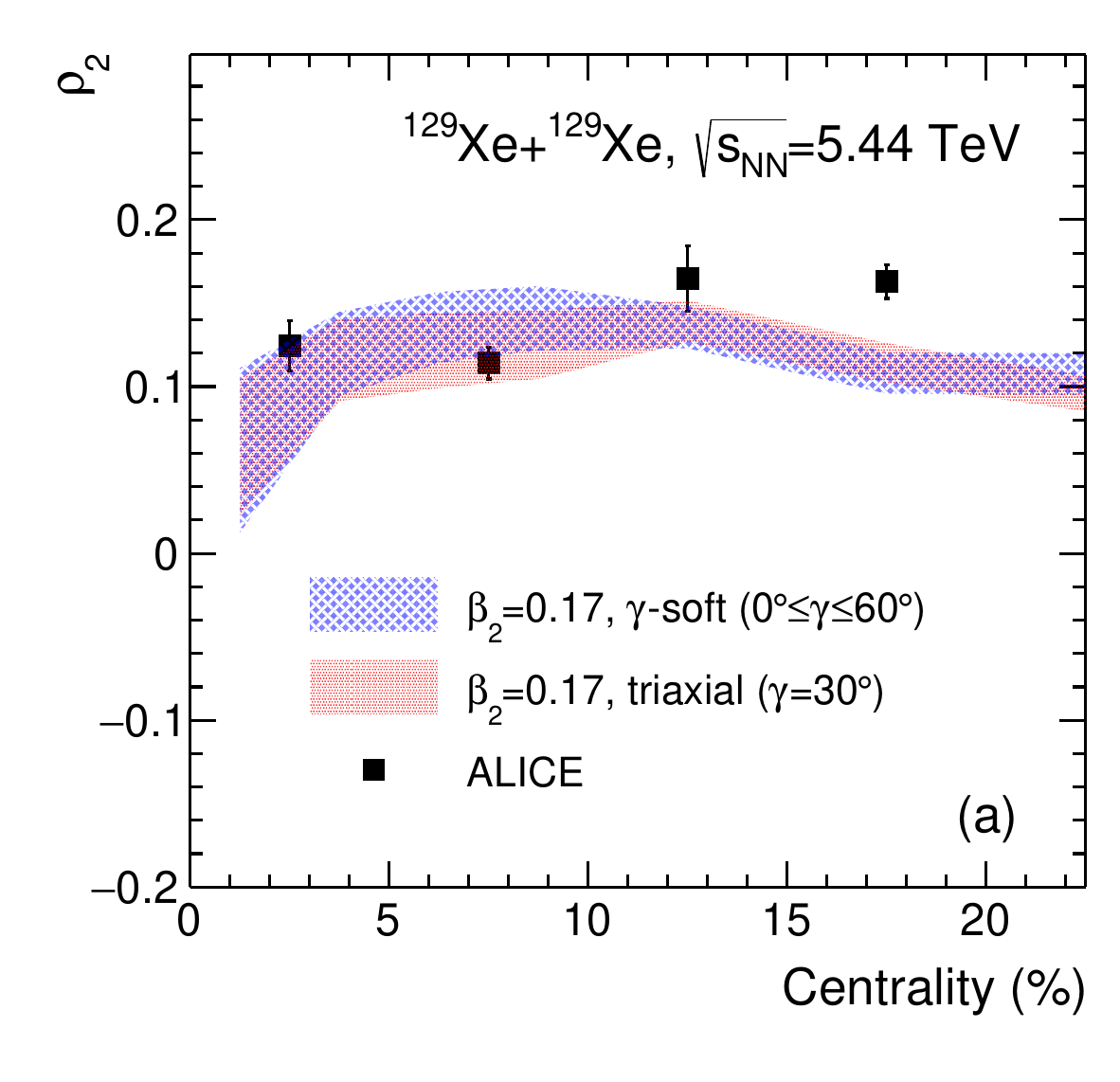}
            \end{minipage}
            \begin{minipage}[t]{0.32\textwidth}
                \includegraphics[scale=0.29]{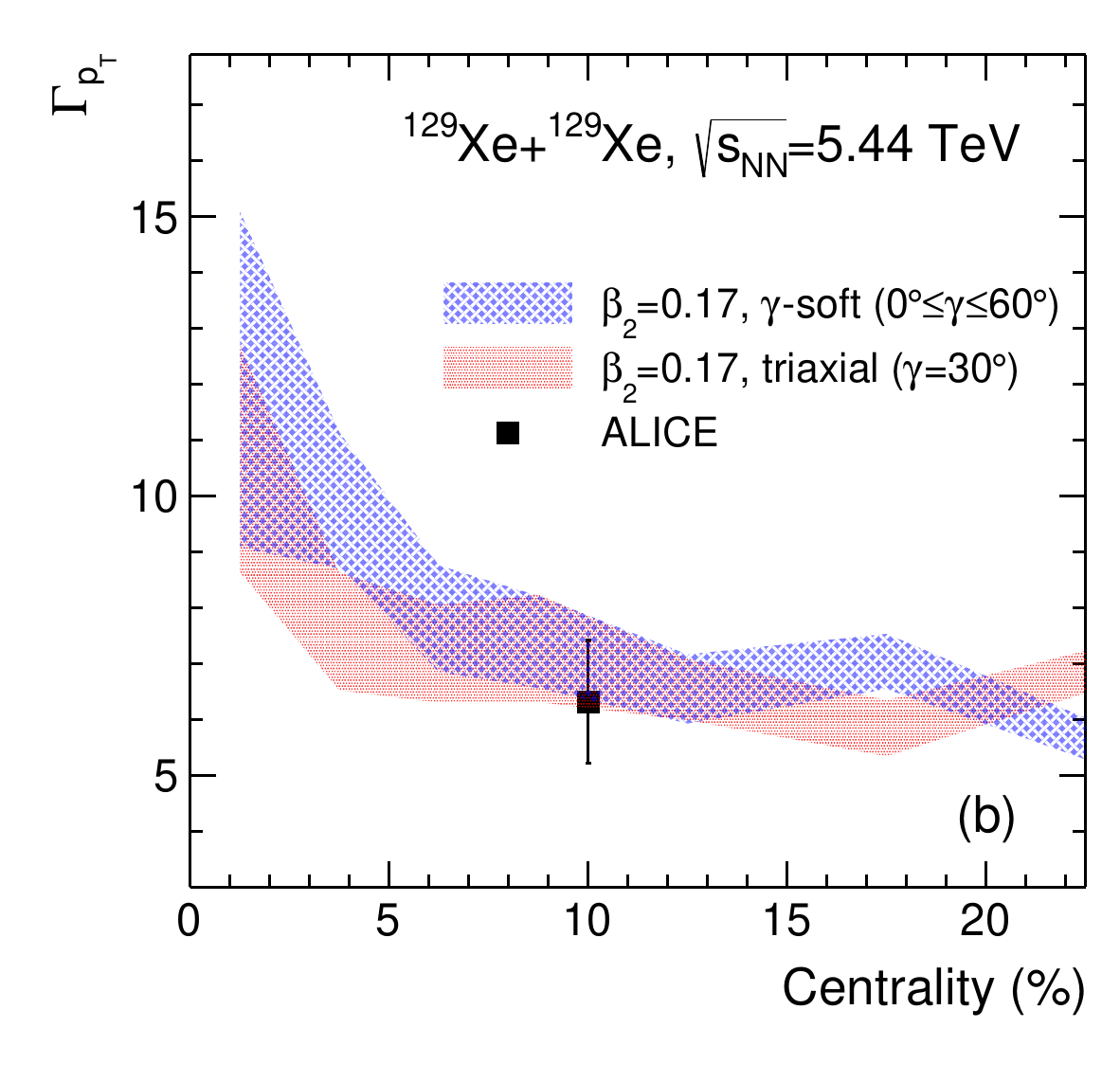}
            \end{minipage}
            \begin{minipage}[t]{0.32\textwidth}
                \includegraphics[scale=0.29]{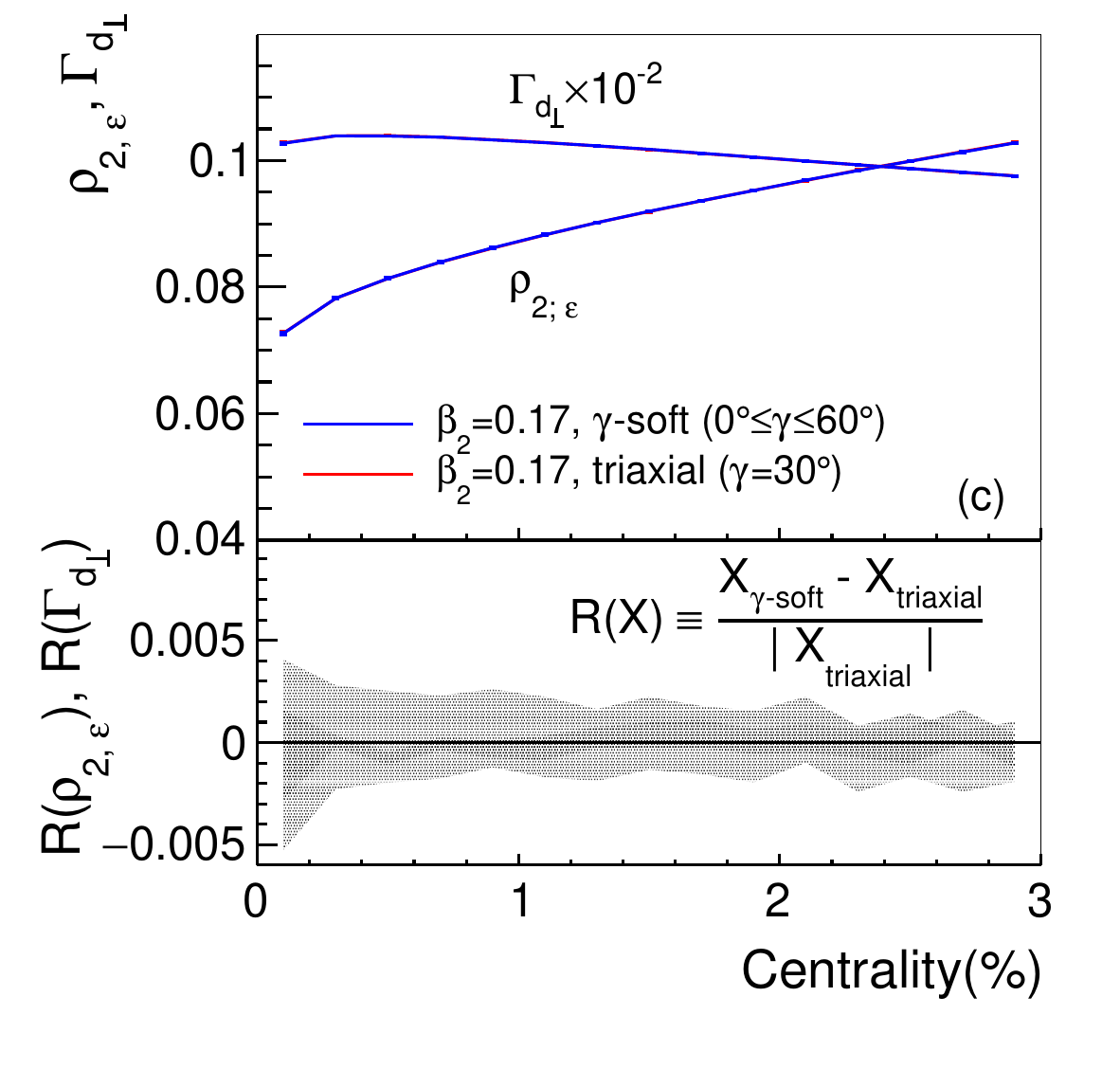}
            \end{minipage}
        \par\end{centering}
        \caption{(a) The pearson coefficient of the \vnpt correlation, $\rho_2$, (b) the intensive skewness of $[p_T]$, $\Gamma_{p_T}$ and (c) The initial stage $\rho_{2;\varepsilon}$ and $\Gamma_{d_\perp}$,  calculated from \ivishnu\ or \trento\ with the $\gamma$-soft and the rigid triaxial deformation of \Xe. The experimental data is taken from ALICE collaboration~\cite{ALICE:2021gxt, ALICE:2023tej}.
        \label{fig:1}}
\end{figure*}

\paragraph{The 3-particle correlations.}
In this section, we focus on the $3$-particle correlations of \XeXe\ collisions to study the deformation of \Xe. Among these 3-particle correlations, two promising observables $\rho_2$ and $\Gamma_{p_T}$ have been used to probe the shape of the colliding nuclei~\cite{Jia:2021qyu, Bally:2021qys,Nielsen:2023znu}. $\rho_2$ measures the correlation between $v_2^{2}$ and $[p_T]$, quantified by the following Pearson correlation coefficient~\cite{Bozek:2016yoj, Schenke:2020uqq, Giacalone:2020dln}:
\begin{align}\label{eq:2}
\rho_2\equiv\frac{\mathrm{cov}(v_2\{2\}^2,[p_T])}{\sqrt{\mathrm{var}(v_2\{2\}^2)}\sqrt{\mathrm{var}([p_T])}}.
\end{align}
where $\mathrm{cov}(v_n\{2\}^2,[p_T])$ denotes the covariance between $v_n\{2\}^2$ measured with the 2-particle correlations and the transverse momentum $[p_T]$. $\mathrm{var}(v_n\{2\}^2)$ and $\mathrm{var}([p_T])$ represent the variance of $v_n\{2\}^2$ and $[p_T]$, respectively.  $\Gamma_{p_T}$ evaluates the intensive skewness of $[p_T]$, defined as ~\cite{Giacalone:2020lbm}:
\begin{align}\label{eq:3}
    \Gamma_{p_T}=\frac{\langle\delta p_{T,i}\delta p_{T,j}\delta p_{T,k}\rangle\langle [p_T]\rangle}{\langle\delta p_{T,i}\delta p_{T,j}\rangle^2},
\end{align}
where $\langle\delta p_{T,i}\delta p_{T,j}\rangle$ and $\langle\delta p_{T,i}\delta p_{T,j}\delta p_{T,k}\rangle$ represent the variance and skewness of the $[p_T]$ distributions, respectively. 
Previous calculations showed different behavior of $\rho_2$ and $\Gamma_{p_T}$ with respect to prolate, triaxial, and oblate deformations of the colliding nuclei~\cite{Bally:2021qys, Nielsen:2023znu}.
Meanwhile, the agreement between the measured $\rho_2$ in \XeXe\ collisions and the initial state model calculations with $\gamma = 27^{\circ}$ was claimed as the first evidence for the rigid triaxial deformation of \Xe~\cite{Bally:2021qys}. However, none of these heavy-ion calculations consider the potential $\gamma$-soft deformation of \Xe\ associated with the second-order shape phase transition along the Xe isotope chain. In Fig.~\ref{fig:1}, we calculate $\rho_2$ and $\Gamma_{p_T}$ in \XeXe\ collisions, using \ivishnu\ simulations with the $\gamma$-soft or the rigid triaxial deformation of \Xe\ modeled in the initial stage. Our calculations with a rigid triaxial deformation of \Xe\ ($\gamma=30^\circ$)  can roughly fit the latest ALICE measurements, as found in~\cite{Bally:2021qys}. Intriguingly, our calculations with the $\gamma$-soft deformation of \Xe\ ($0^\circ\le\gamma\le60^\circ$) can also simultaneously describe the measured $\rho_2$ and $\Gamma_{p_T}$, which are fully consistent with the calculations using the rigid triaxial deformation of \Xe~\footnote{Using the same kinetic cut and centrality determination as the ATLAS collaboration, we can also reproduce the measured $\rho_2$ from ATLAS.}.

Fig.~\ref{fig:1} (c) plots the corresponding initial state correlations  $\rho_{2;\varepsilon}$ and $\Gamma_{d_\perp}$, defined as:
\begin{align}
\rho_{2;\varepsilon}&=\frac{\mathrm{cov}(\varepsilon_2\{2\}^2,d_\perp)}{\sqrt{\mathrm{var}(\varepsilon_2\{2\}^2)}\sqrt{\mathrm{var}(d_\perp)}},\\
\Gamma_{d_\perp}&=\frac{\langle(\delta d_{\perp})^3\rangle\langle d_\perp\rangle}{\langle(\delta d_\perp)^2\rangle^2},
\end{align}
These two definitions are similar to the $\rho_{2}$ and $\Gamma_{p_T}$ above, but replace the elliptic flow $v_2$ and the transverse momentum $p_T$ with the  eccentricity $\varepsilon_n=-\{r^ne^{in\phi}\}/\{r^n\}$ and the initial energy per particle $d_\perp\equiv E/S$~\cite{Giacalone:2020dln}, using the  hydrodynamic linear response relation~\cite{Schenke:2020uqq, Niemi:2015qia, Jia:2021qyu, Dimri:2023wup}. Here, $\delta d_\perp$ represents the fluctuations of the initial gradient $d_\perp$.  Fig.~\ref{fig:1} (c) shows that the curves calculated with $\gamma$-soft deformation almost overlap with the curves with rigid triaxial deformation, further demonstrating that the measured 3-particle correlations $\rho_2$ and  $\Gamma_{p_T}$ have no discriminating power with respect to the different $\gamma$-structure of $^{129}$Xe. As pointed out by the simple liquid-drop model~\cite{Jia:2021qyu}, the $\gamma$ dependence of both $\rho_{2;\varepsilon}$ and $\Gamma_{d_\perp}$ are sensitive only to $\langle \cos (3 \gamma) \rangle$. Thus, these two observables in heavy ion collisions can not distinguish between the two scenarios of \Xe\ deformation as long as they produce approximately the same $\langle \cos (3 \gamma) \rangle$.
Considering the 3-particle correlations in Fig.~\ref{fig:1}, a firm conclusion on the rigid triaxial deformation or the $\gamma$-soft deformation of \Xe\ from the $\rho_2$ and $\Gamma_{p_T}$ measurements in \XeXe\ collisions at the LHC remains elusive.

\begin{figure*}[!hbt]
        \begin{centering}
            \begin{minipage}[t]{0.4\textwidth}
                \includegraphics[scale=0.35]{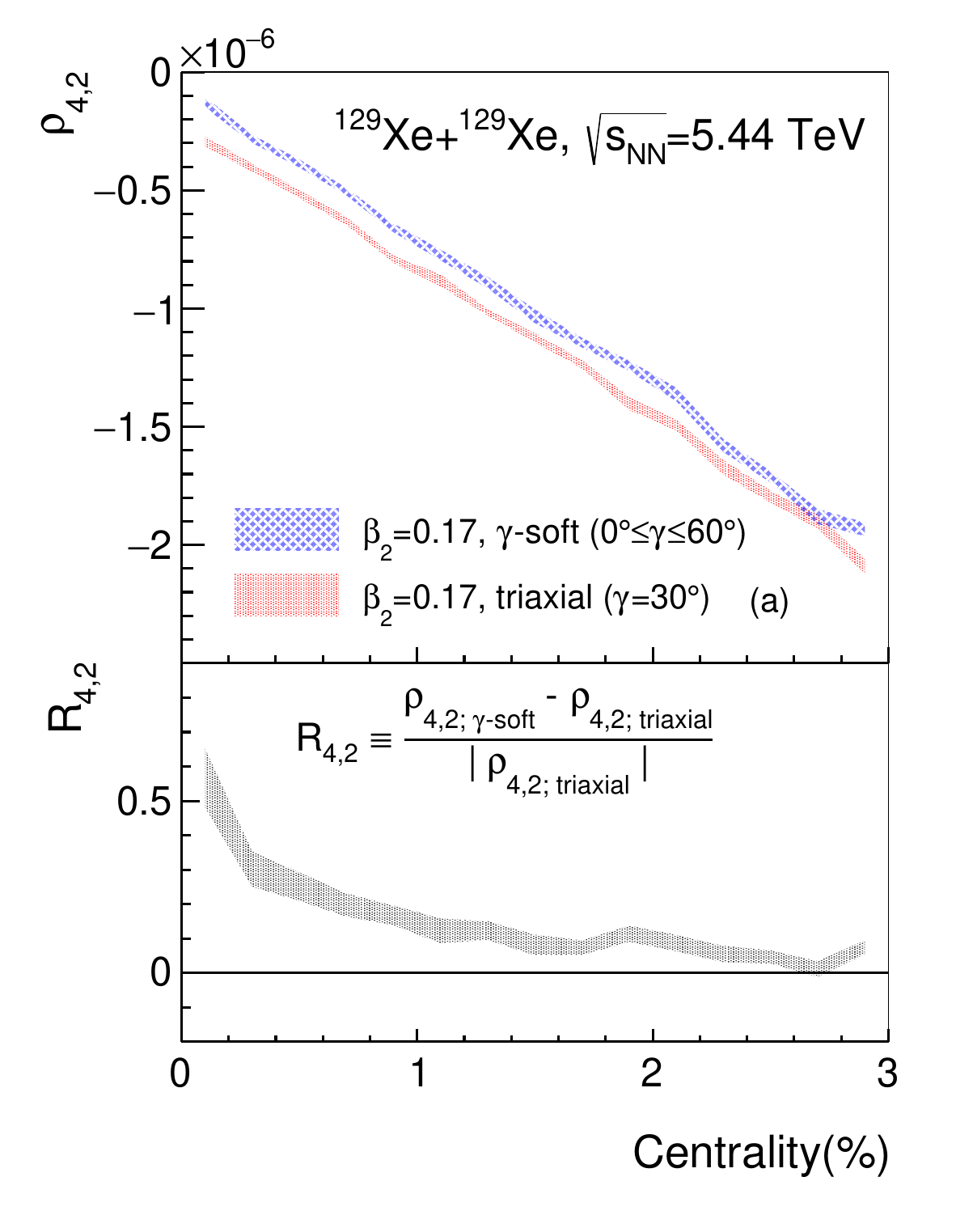}
            \end{minipage}
            \begin{minipage}[t]{0.4\textwidth}
                \includegraphics[scale=0.35]{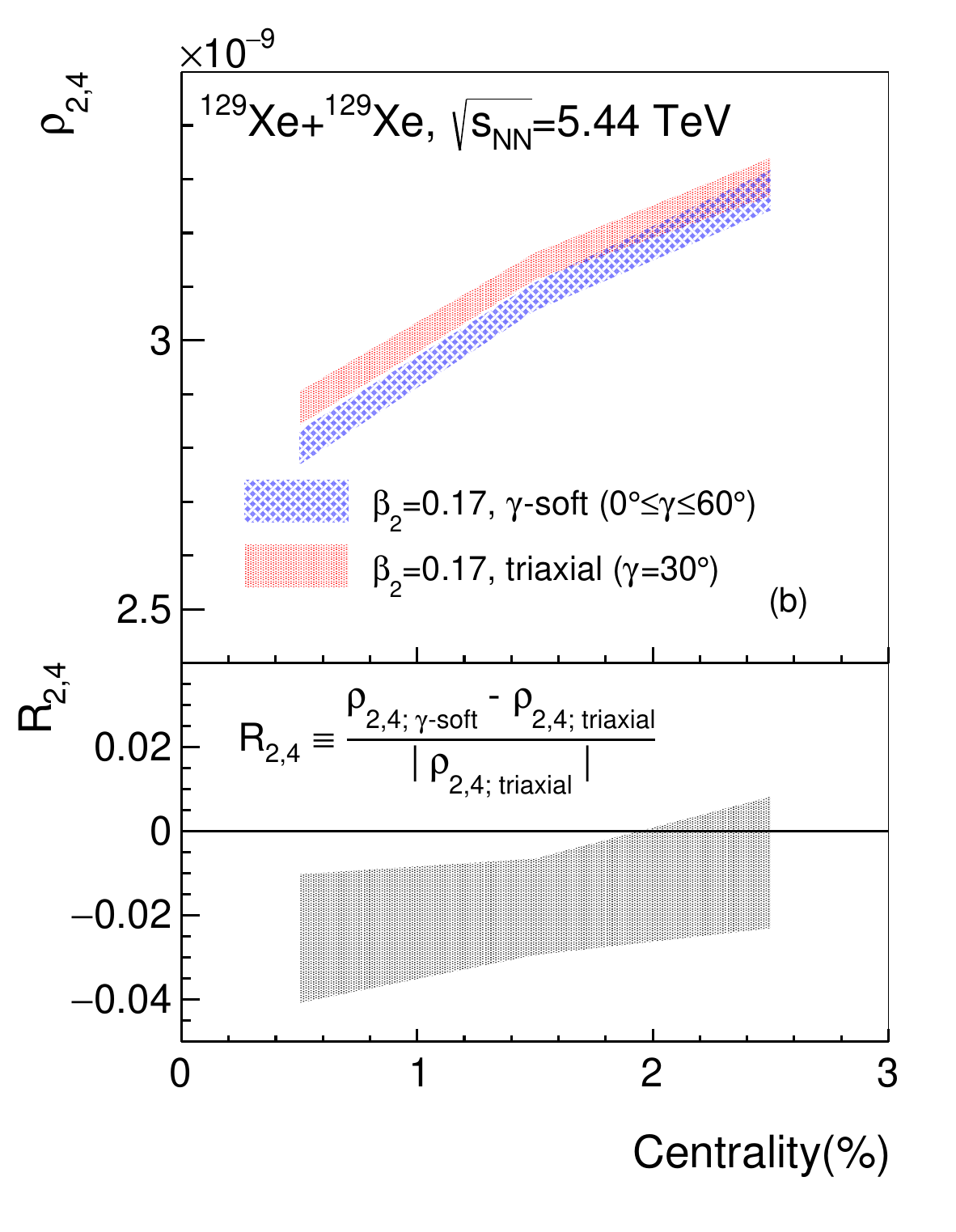}
            \end{minipage}
        \par\end{centering}
        \caption{(a) $\rho_{4,2}$ and (b) $\rho_{2,4}$ calculated from \ivishnu\ with the $\gamma$-soft or rigid triaxial deformation of \Xe.
         The ratio of $\rho_{4,2}$ and $\rho_{2,4}$ between two different deformations of \Xe\ are shown at the bottom.
        \label{fig:2}}
\end{figure*}

\paragraph{The 6-particle correlations.}
To distinguish between the $\gamma$-soft and the rigid triaxial deformation of \Xe\ in heavy ion collisions, we now construct new correlators, the 6-particle correlations, which have a unique sensitivity to $\langle\cos(3n\gamma)\rangle$ with $n\ge2$. Predicting the higher-order correlations from hydrodynamics requires a high statistical run and consumes a large amount of computational resources. Considering the linear response of hydrodynamics, which maps the final state flow harmonics to the initial state eccentricities at lower order, we focus on constructing and studying the 6-particle correlations from the initial state. Here we construct two new 6-particle correlators of the initial state:

\begin{widetext}
    \begin{align}
    \rho_{4,2}\equiv\left(\frac{\langle \varepsilon_2^4\delta \dperp^2\rangle}{\langle \varepsilon_2^4\rangle\langle \dperp\rangle^2}\right)_c &\equiv \frac{1}{\langle \varepsilon_2^4\rangle\langle \dperp\rangle^2}\left[\langle \varepsilon_2^4\delta \dperp^2\rangle+4\langle \varepsilon_2^2\rangle^{2}\langle\delta\dperp^2\rangle-\langle \varepsilon_2^4\rangle\langle\delta\dperp^2\rangle-4\langle \varepsilon_2^2\rangle\langle \varepsilon_2^2\delta\dperp^2\rangle-4\langle \varepsilon_2^2\delta\dperp\rangle^2\right]\\
    \rho_{2,4}\equiv\left(\frac{\langle \varepsilon_2^2\delta \dperp^4\rangle}{\langle \varepsilon_2^2\rangle\langle\dperp\rangle^4}\right)_c &\equiv \frac{1}{\langle \varepsilon_2^2\rangle\langle \dperp\rangle^4}\left[\langle \varepsilon_2^2\delta \dperp^4\rangle-6\langle \varepsilon_2^2\delta\dperp^2\rangle\langle\delta\dperp^2\rangle-4\langle \varepsilon_2^2\delta\dperp\rangle\langle\delta\dperp^3\rangle-\langle \varepsilon_2^2\rangle\langle\delta\dperp^4\rangle+6\langle \varepsilon_2^2\rangle\left(\langle\delta\dperp^2\rangle\right)\right].
    \end{align}
\end{widetext}
Using the estimates from the liquid-drop model~\cite{Jia:2021qyu, Dimri:2023wup}, we express for the two quantities with the following $\beta_2$ and $\gamma$ dependence:
\begin{align}\label{eq:4}
    \begin{split}
    \mean{\varepsilon_2^4}\rho_{4,2} =A\beta_2^6(53+16\mean{\cos(6\gamma)})+f_{4,2}(\beta_2^6, \mean{\cos(3\gamma)}),\\
    \mean{\varepsilon_2^2}\rho_{2,4} =\frac{A}{16}\beta_2^6(43-14\mean{\cos(6\gamma)})+f_{2,4}(\beta_2^6, \mean{\cos(3\gamma)}),
    \end{split}
\end{align}
where $A\equiv\frac{75}{1025024\pi^3}$, $f_{4,2}$ and $f_{2,4}$ are the functions of $\beta_2^6$ and $\mean{\cos(3\gamma)}$. The above expressions are obtained with the assumptions of no $\beta_2$ fluctuation and independent rotation of the two colliding nuclei. For our current $\gamma$ configurations for either the $\gamma$-soft or the rigid triaxial deformation of \Xe, $\langle\cos3\gamma\rangle=0$, which means that $f_{4,2}$ and $f_{2,4}$ terms are negligible.

Fig.~\ref{fig:2} plots the centrality dependent $\rho_{4,2}$ and $\rho_{2,4}$, calculated from \trento\ model with the same initial state parameters as used in the full iEBE-VISHNU simulations.  
Fig.~\ref{fig:2}(a) shows that, compared to the case of triaxially deformed \Xe, the $\gamma$-soft deformation increases $\rho_{4,2}$. As derived in Eq.\eqref{eq:4}, the associated term $\langle\cos(6\gamma)\rangle=-1$ for the rigid triaxial deformation with $\gamma=30^\circ$, and $\langle\cos(6\gamma)\rangle=0$ for the $\gamma$-soft deformation with $0^\circ\le\gamma\le60^\circ$. This explains the increasing trend of $R(\rho_{4,2})$ for the $\gamma$-soft case. In the bottom panel of Fig.~\ref{fig:2}(a), we also plot the ratio between the two deformations of \Xe, $R(\rho_{4,2})$, defined as  $R(X)=\frac{X_{\text{$\gamma$-soft}}-X_{\text{triaxial}}}{|X_{\text{triaxial}}|}$.
With $\beta_2$ set to 0.17, the effect of $\gamma$-soft can be observed for the selected centrality range, with the enhancement of $R(\rho_{4,2})$ increasing to $50\%$ around 0.2\% ultra-central collisions.  Due to the $\beta_2^6$ dependence shown in Eq.\eqref{eq:4}, the $\gamma$-soft effect can be significantly enhanced with a larger value of $\beta_2$. Calculations from the nuclear structure predict that the quadrupole deformation of \Xe\ is about $\beta_2 = 0.17\sim0.20$~\cite{Raman:2001nnq, Elekes:2015qje}. We find that with $\beta_2=0.2$, the enhancement of $R(\rho_{4,2})$ can increase to $100\%$ around 0.2\% ultra-central collisions. This shows the discriminatory power of $\rho_{4,2}$ to identify the $\gamma$-soft and the triaxial deformation of \Xe\ in ultra-central \XeXe\ collisions.

Fig.~\ref{fig:2}(b) plots the $\rho_{2,4}$ for the selected centrality range. 
A small suppression within 2\% can be observed after changing from the rigid triaxial deformation to the $\gamma$-soft deformation in the colliding nuclei \Xe. The $\gamma$-soft effect is much less significant compared to the $\rho_{4,2}$ results shown in Fig.~\ref{fig:2}(a). This is qualitatively consistent with the expectations of the liquid-drop model with the associated  $\langle\cos(6\gamma)\rangle$ term in Eq.\eqref{eq:4}, which predicts an order of magnitude weaker $\gamma$-soft effect for $\rho_{2,4}$. Despite the large statistical uncertainties of the $\rho_{2,4}$ results, a clear distinction between the $\gamma$-soft and the rigid triaxial deformation are already visible at $\beta_2=0.17$. The discriminating power of $\rho_{2,4}$ will be further enhanced for larger deformations, e.g., $\beta_2=0.20$. Therefore, we propose to simultaneously study the associated final state correlators for $\rho_{2,4}$ and $\rho_{4,2}$ in \XeXe\ collisions at the LHC, together with a comparison with the calculations from the state-of-the-art hydrodynamic model. This will provide a novel opportunity to discover the second-order shape phase transition associated with the $\gamma$-soft deformation of \Xe\ or to confirm the rigid triaxial deformation of  \Xe\ via the observables in relativistic heavy-ion collisions at the TeV energy scale.

\paragraph{Summary.}
The shape phase transition of finite nuclei for certain isotope or isotone chains is a fascinating phenomenon in nuclear physics. Calculations and measurements in nuclear structure physics suggest that a second-order shape phase transition occurs for the Xe isotopes near $^{128-130}$Xe. In this letter, we investigate the $\gamma$-soft deformation of \Xe,\ associated with the possible second-order shape phase transition with the constructed novel correlators in ultra-relativistic \XeXe\ collisions. Using the state-of-the-art hybrid model \ivishnu\ within the hydrodynamic framework, we study the multi-particle transverse momentum correlations and the correlations between the mean transverse momentum and the elliptic flow. Our calculations show that these measured $3$-particle correlators, $\rho_2$ and $\Gamma_{p_T}$, which were previously interpreted as the evidence for the rigid triaxial deformation of \Xe, can also be well explained by the $\gamma$-soft deformation of \Xe. These two correlators are effective in probing the mean value of the $\gamma$ distribution of the colliding nuclei, which is insensitive to the $\gamma$-soft or the rigid triaxial deformation of \Xe. 

In order to distinguish between the $\gamma$-soft and the rigid triaxial deformation of \Xe\ in ultra-relativistic \XeXe\ collisions, we construct two new correlators, $\rho_{4,2}$ and $\rho_{2,4}$, which carry non-trivial higher-order correlations, and demonstrate the discernible sensitivity of $\rho_{4,2}$ to the different deformations of \Xe.
By performing precise experimental measurements for higher-order correlators associated with $\rho_{4,2}$ and $\rho_{2,4}$,
together with sophisticated hydrodynamic model calculations using the Bayesian analysis, it is possible to constrain the deformation of \Xe\ in \XeXe\ collisions at the LHC. The present study also provides a first but crucial step towards exploring the second-order shape phase transition of finite nuclei in ultra-relativistic heavy-ion collisions.  We expect that the current and future research will expand the existing scientific scope of high-energy nucleus-nucleus collisions at RHIC and the LHC and complement the discovery of novel phenomena in low-energy nuclear physics.

\hspace*{\fill}

\paragraph{Acknowledgement.} We thank Jens J\o rgen Gaardh\o je, Fuqiang Wang, Jiangming Yao, and Yu Zhang for the constructive discussions and suggestions.
This work is supported in part by the National Natural Science Foundation of China with grant Nos.12247107, 12075007, 12175007, 12275082, 12035006, 12075085, the European Union (ERC, Initial Conditions), VILLUM FONDEN with grant no. 00025462, and Danmarks Frie Forskningsfond (Independent Research Fund Denmark).

\bibliography{apssamp.bib}
\end{document}